\documentclass[10pt,letterpaper]{article}

\usepackage{ccn}
\usepackage{pslatex}
\usepackage{apacite}
\usepackage{color}
\usepackage{soul}
\usepackage{amsmath}
\usepackage{bm}
\usepackage{graphicx}
\usepackage{verbatim}

\title{Modulation of early visual processing alleviates capacity limits \\in solving multiple tasks}
 
\author{{\large \bf Sushrut Thorat\thanks{Equal contribution} \qquad Giacomo Aldegheri\footnotemark[1] \qquad Marcel A. J. van Gerven \qquad Marius V. Peelen} \\
s.thorat@donders.ru.nl, g.aldegheri@donders.ru.nl, m.vangerven@donders.ru.nl, m.peelen@donders.ru.nl\\
  Donders Institute for Brain, Cognition and Behaviour, Radboud University}

\begin{document}

\maketitle

\section{Abstract}
{
\bf

In daily life situations, we have to perform multiple tasks given a visual stimulus, which requires task-relevant information to be transmitted through our visual system. When it is not possible to transmit all the possibly relevant information to higher layers, due to a bottleneck, task-based modulation of early visual processing might be necessary.
In this work, we report how the effectiveness of modulating the early processing stage of an artificial neural network depends on the information bottleneck faced by the network. The bottleneck is quantified by the number of tasks the network has to perform and the neural capacity of the later stage of the network. The effectiveness is gauged by the performance on multiple object detection tasks, where the network is trained with a recent multi-task optimisation scheme. By associating neural modulations with task-based \textit{switching} of the state of the network and characterising when such switching is helpful in early processing, our results provide a functional perspective towards understanding why task-based modulation of early neural processes might be observed in the primate visual cortex\footnote{The code to train and analyse the networks mentioned here can be found at - https://github.com/novelmartis/early-vs-late-multi-task}\footnote{Accepted into the 2019 Conference on Cognitive Computational Neuroscience: \url{https://doi.org/10.32470/CCN.2019.1229-0}}.
}
\begin{quote}
\small
\textbf{Keywords:} 
neural modulation, multi-task learning, early visual cortex, attention, perception, capacity limits
\end{quote}

\section{Introduction}
Humans and other animals have to perform multiple tasks given a visual stimulus. For example, seeing a face, we may have to say whether it is happy or sad, or recognise its identity. For each of these tasks, a subset of all the features of the face are useful. In principle, it could be possible for a visual system to extract all of the features necessary to solve all possible tasks, and then select the relevant information from this rich representation downstream.
However, as the number of tasks increases, a network with a limited capacity may not be able to extract all of the potentially relevant features (an information bottleneck is manifest), requiring the information that is extracted from the stimulus in the early processing stages to change according to the task.

Several studies in neuroscience have found evidence for such task-dependent modulations of sensory processing in the primate visual system, including at the early levels~\cite{carrasco2011visual,maunsell2006feature,gilbert2013top}.
For example, human neuroimaging studies have shown that attending to a stimulus could lead to an increase in the accuracy with which its task-relevant features could be decoded by a classifier in early visual areas \cite{jehee2011attention}, and neurophysiological experiments in nonhuman primates have shown that the stimulus selectivity of neurons in primary visual cortex was dependent on the task the monkeys had to perform \cite{gilbert2013top}.

Despite the observation of such modulations of early visual processing, it is not clear whether they are causally necessary for performing better on the corresponding tasks.
This question has been addressed by deploying biologically-inspired task-based modulations on computational models. \citeA{lindsay2018biological} showed that task-based modulation deployed on multiple stages of a convolutional neural network improves performance on challenging object classification tasks. Other recent work~\cite{thorat2018functional,rosenfeld2018priming} has also shown that task-based modulation of early visual processing aids in object detection and segmentation in addition to the task-based modulation of late processing. However, the conditions under which early modulation can be beneficial in performing multiple tasks have not been systematically investigated.

In the present work, we assessed the effectiveness of task-based modulation of early visual processing as a function of an information bottleneck in a neural network, quantified by the number of tasks the network had to execute and the neural capacity of the network. To do so, we trained networks to, given an image, provide an answer conditioned on the cued task. Every task required detecting the presence of the corresponding object in the image. The networks were trained according to a recent framework proposed in the field of continual learning \cite{cheung2019superposition}, which helps them execute multiple tasks by switching their state given a task cue, in order to transmit relevant information through the network.
In this work, to quantify the effectiveness of task-based modulation of early neural processing, we measured the increase in performance provided by modulating early neural processing in addition to modulating the late neural processing in the networks.

\section{Methods}

\subsection{Task and system description}

In a multi-task setting, object detection can be thought of as solving one of a set of possible binary classification (one object versus the rest) problems. Given an image and a task cue indicating the identity of the object to be detected, a network had to output if the object in the image matched the task cue. 

We used MNIST~\cite{lecun1998gradient} digits and their permutations as objects~\cite{kirkpatrick2017overcoming}. The original MNIST dataset has $28\times 28\,$px$^{2}$ images of $10$ digits. Each permuted version consists of images of those $10$ digits, whose pixels undergo a given permutation, creating $10$ new objects. We varied the number of permutations used ($10$, $25$, and $50$) to modulate the number of tasks the networks had to perform (which are $10$ times the number of permutations).

We considered a multi-layer perceptron with rectified linear units (ReLU), which had one hidden layer between the input (image) and the binary output. The number of neurons in the hidden layer were variable ($32$, $64$, and $128$) and determined the neural capacity of the late stage of the network.

\subsection{Task-based modulation and its function}

Modelling biological neurons as perceptrons~\cite{rosenblatt1957perceptron}, task-based modulations have been shown to affect the effective biases and gains of the neurons~\cite{maunsell2006feature,boynton2009framework,ling2009spatial}. The nature of modulation - which neurons to modulate and how - is under debate~\cite{boynton2009framework,thorat2018functional}. We adapted these findings by introducing task-based modulation into our networks via the biases of the perceptrons and the gains of their ReLU activation functions. The modulations were then trained end-to-end with the rest of the network.

Given a particular task, the task cue is a one-hot encoding of the relevant object. Task-based modulation is mediated through bias and gain modulation in the following manner.
\begin{align}
    x_{n} &= [W_{n}(g_{n-1}\circ x_{n-1}) + b_n]_+ \\
    g_{n} &= G_{n}c,\,\,\, b_{n} = B_{n}c
\end{align}
where the transformation between layers $n-1$ and $n$ ($L_{n-1}\rightarrow L_n$) is modulated by changing the slope of the ReLU activation function (gains, $g_{n-1}$) in $L_{n-1}$ and the biases ($b_n$) to the perceptrons in $L_n$; $x_{n}$ are the pre-gain activations of the perceptrons in $L_n$, $W_{n}$ is the task-independent transformation matrix between $L_{n-1}$ and $L_n$, $G_{n}$ and $B_{n}$ map the task cue $c$ (one-hot encoding of the relevant object $k$) to the gain and bias modulations of the perceptrons in $L_n$ respectively, and $\circ$ refers to element-wise multiplication.

\begin{figure}[!t]
\centering
\includegraphics[width=0.4\textwidth]{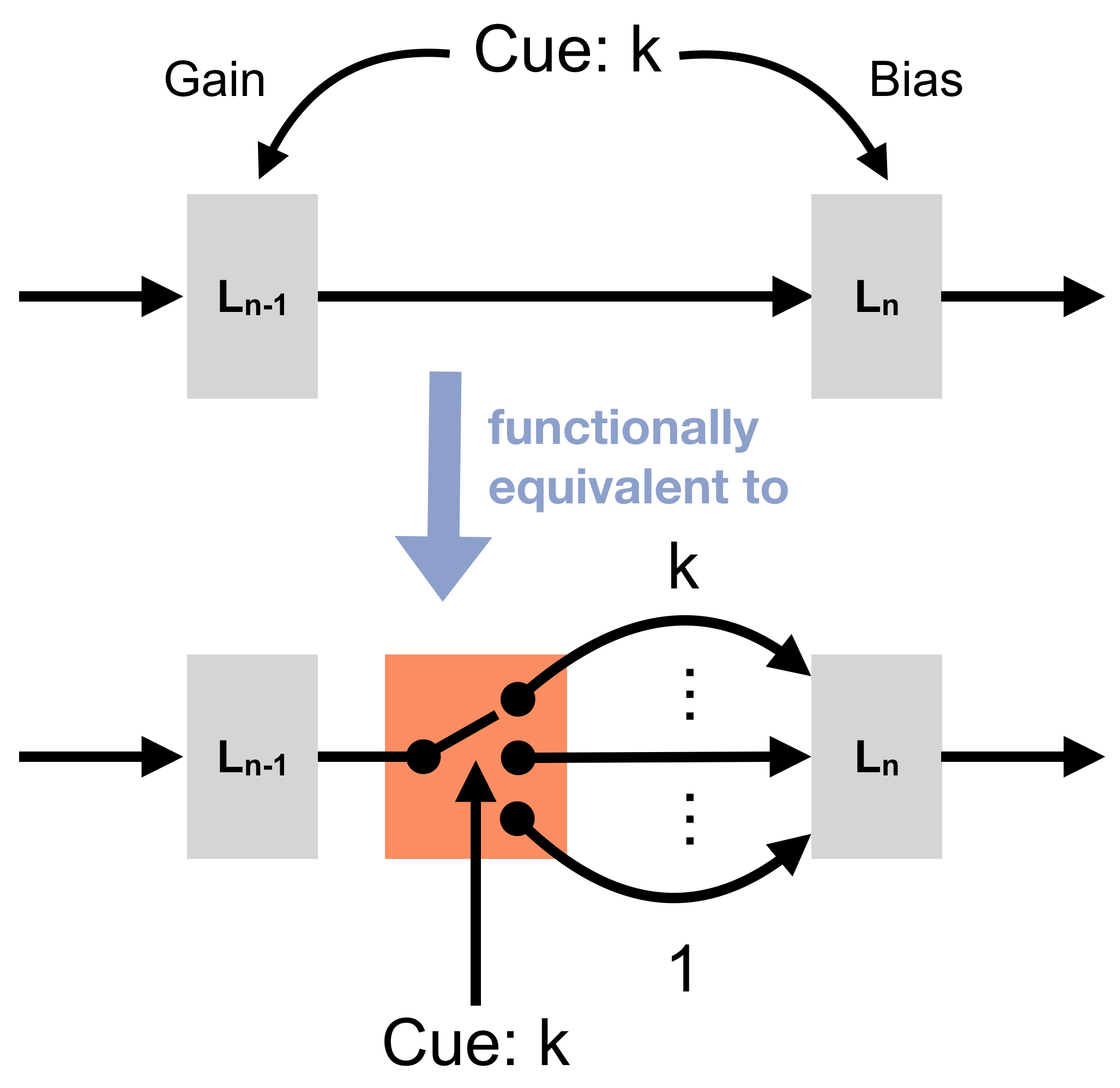}
\caption{The effect of bias and gain modulation on the transformations in the network. Modulating the gains and biases is \textit{functionally} equivalent to \textit{switching} the transformation being performed to one suited for the relevant task. Such an example of switching is visualised in the figure. Given a task cue corresponding to object $k$, corresponding gain and bias modulations are applied, which results in the $L_{n-1}\rightarrow L_n$ transformation being switched into one that transmits feature information required to detect the presence or absence of object $k$.}
\label{fig:fig2}
\end{figure}

Given a task $k$, modulating the gains of the pre-synaptic perceptrons (in $L_{n-1}$) and the biases of the post-synaptic perceptrons (in $L_n$) transforms the information transformation between $L_{n-1}$ and $L_n$. This allows for the transmission of information required to perform task $k$, while ignoring the information required to perform the other tasks, as formalised in \citeA{cheung2019superposition}. This transformation can also be thought of as the network \textit{switching} its state to selectively transmit task-relevant information downstream (see Figure~\ref{fig:fig2}). The conditions - the nature of these modulations and the neural capacity of the network - under which the network can switch between a given number of tasks, are preliminarily described in \citeA{cheung2019superposition}.

Here, for every relevant layer $L_n$, $W_{n}$, $B_{n}$, and $G_{n}$ were jointly learned for the given number of tasks.

\subsection{Evaluation metric and expected trends}

The effectiveness of early neural modulation was quantified by the average absolute increase in detection performance across all the tasks when modulations were implemented on both the transformations $L_1\rightarrow L_2$ and $L_2\rightarrow L_3$ ($L_1$ corresponds to the input layer and $L_3$ to the output layer) as opposed to when the modulations were trained on the transformation $L_2\rightarrow L_3$ only.

We expected the effectiveness of task-based early neural modulation to be directly proportional to the number of neurons in $L_2$ and inversely proportional to the number of tasks (permuted MNIST sets used).

\subsection{Neural network training details}

All the networks were trained with adaptive stochastic gradient descent with backpropagation through the ADAM optimiser~\cite{kingma2014adam} with the default settings in TensorFlow (v1.4.0) and $\alpha = 10^{-5}$. We used a batch size of $100$. Half of each batch contained randomly selected images of randomly selected tasks where the cued object was present, and half where the cued object was not present. These images were taken from the MNIST training set and its corresponding permutations. The images were augmented by adding small translations and noise. We trained each network with $10^7$ such batches. The relevant metrics discussed in the previous section are computed at the end of training over a batch of size $10^5$ created from the MNIST test set and its corresponding permutations.

\section{Results}

\begin{figure}[!t]
\centering
\includegraphics[width=0.48\textwidth]{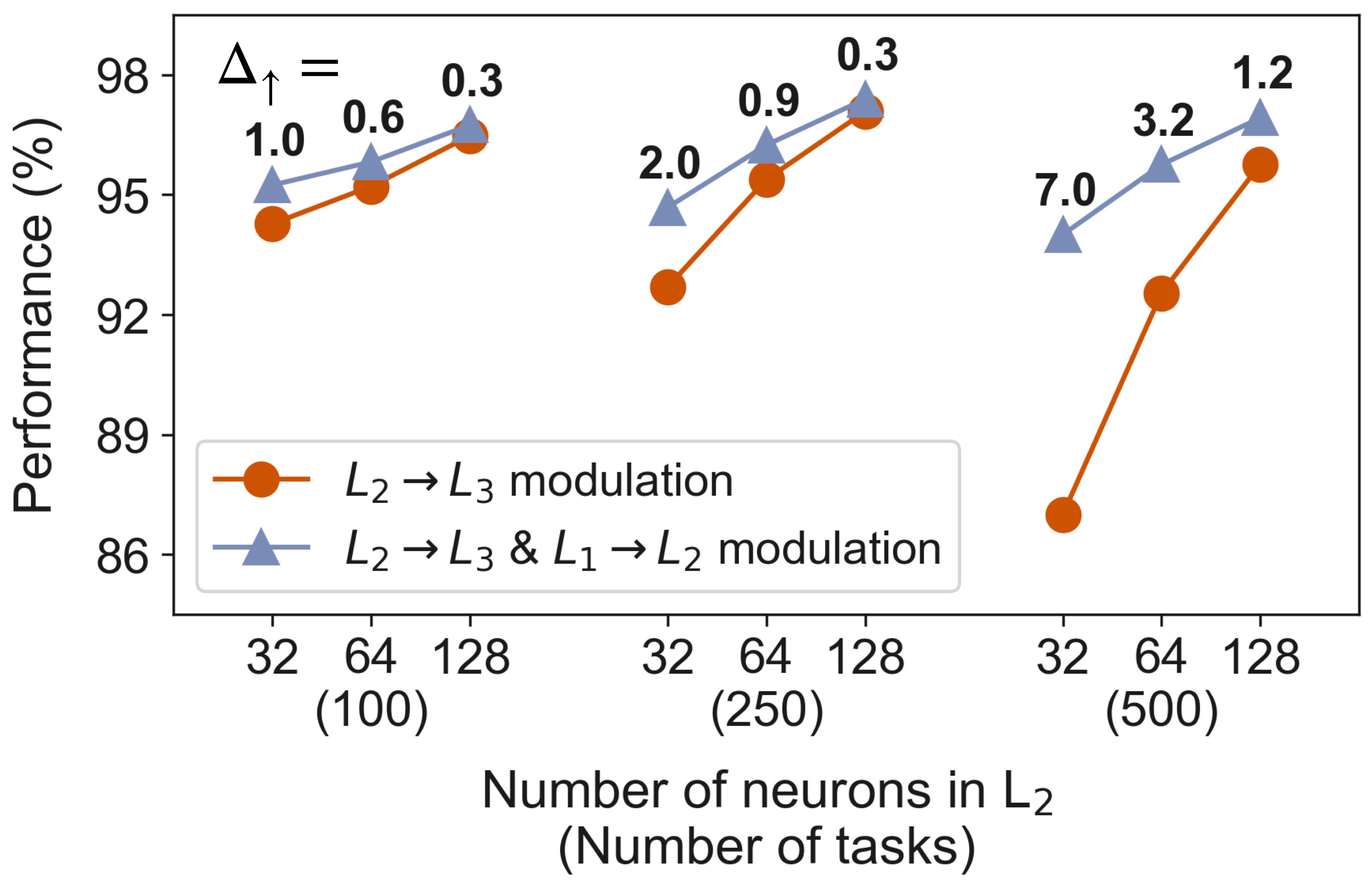}
\caption{The effectiveness of task-based modulation (quantified by the performance boost, $\Delta_{\uparrow}$) of early neural processing ($L_1\rightarrow L_2$) as a function of the number of neurons in $L_2$ and the number of tasks the network has to perform. The performance boost was inversely proportional to the number of neurons in $L_2$ and directly proportional to the number of tasks the network had to perform. The absolute performance profiles given either the modulation of $L_2\rightarrow L_3$ only or the joint modulation of $L_2\rightarrow L_3$ and $L_1\rightarrow L_2$ are also shown. (See Appendix for further substantiation of these results)}
\label{fig:fig3}
\end{figure}

We first analysed the detection performance of the network with only $L_2\rightarrow L_3$ modulation. The network performance as a function of the number of neurons in $L_2$ and the number of detection tasks the network had to perform is shown in Figure~\ref{fig:fig3} (red circles). The network performance increased with an increase in the number of neurons in $L_2$, as the neural capacity increased. The performance decreased with an increase in the number of tasks to be performed, as the representational capacity of the network for any one task was reduced. A network with as little as $32$ neurons in its hidden layer was able to switch between as many as $500$ detection tasks, while keeping the average detection performance across all the tasks as high as $87\%$, thus replicating the success of the multi-task learning framework proposed by \citeA{cheung2019superposition}. 

To assess the dependence of the effectiveness of task-based modulation of early neural processing ($L_1\rightarrow L_2$) on the bottleneck in the network, we analysed the boost in average detection performance when task-based modulation of $L_1\rightarrow L_2$ was deployed in addition to task-based modulation of $L_2\rightarrow L_3$, as a function of the number of neurons in $L_2$ and the number of detection tasks the network had to perform. The resulting boosts are shown in Figure~\ref{fig:fig3} ($\Delta_{\uparrow}$ quantification). The performance boost increased as the number of neurons in $L_2$ decreased, and as the number of tasks the network has to perform increased. This confirms the hypothesis that task-based modulation of early neural processing is essential when an information bottleneck exists in a subsequent processing stage (see Appendix for further analyses elucidating these results).

\subsection{The contribution of bias and gain modulation}

Gain, but not so much bias, modulation of neural responses has been observed in experiments investigating feature-based attention in the monkey/human brain~\cite{maunsell2006feature,boynton2009framework}. We assessed how the two contributed to the overall modulation of the transformations in the network.

We selectively turned off the bias or gain modulation for all the variants of the network that were trained. The average detection performance decreased by $43.0\pm 2.0\%$ when gain modulation was turned off, and by $3.9\pm 0.9\%$ when bias modulation was turned off, suggesting that in our framework, when jointly deployed, gain modulation is more important than bias modulation in switching the state of the network to be able to perform the desired task well. 

We also trained a network with $32$ neurons in $L_2$, on $25$ permutations of MNIST, with gain-only or bias-only modulations of both the $L_1\rightarrow L_2$ and $L_2\rightarrow L_3$ transformations. When the gain and bias modulations were jointly trained, the network performance was $94.7\%$. With gain-only modulation, the performance was $94.8\%$, and with bias-only modulation the performance was $90.9\%$. As the performance when only bias modulation was deployed was much higher than chance ($50\%$), we can conclude that bias modulation alone can also lead to efficient task-switching. When the bias and gain modulations are jointly trained, gain might take over as it multiplicatively impacts responses, and therefore has higher gradients during training, as opposed to the additive impact of bias.

\section{Discussion}

Adding to the discussion about the functional role of task-based modulation of early neural processing, in this work we have shown that modulating the early layer of an artificial neural network in a task-dependent manner can boost performance, beyond just modulating the late layer, in a multi-task learning scenario in which a network contains an information bottleneck, either due to a large number of tasks to be performed or to a small number of units in the late layer.

Adapting a formalism proposed by \citeA{cheung2019superposition}, we showed how bias and gain modulation, two prevalent neuronal implementations of top-down modulation in the brain, could functionally lead to switching the state of a network to perform transformations effective for the task at hand. 
While task-dependent computations are widespread in higher-level areas of the primate brain, such as prefrontal cortex~\cite{mante2013context}, it is not clear to what extent sensory streams (which perform early visual processing) can also be seen as switching their state according to the current task (although see \citeA{gilbert2013top} for a proposal), and what the functional relevance of doing so would be.
Here we show how, in principle, this switching could be computationally advantageous when it is not possible to send the information required for all tasks to higher layers, which might well be the case in the complex environments that humans and other animals are able to navigate.

To further investigate the relevance of our findings to biological visual systems, in follow-up work we intend to deploy our modulation scheme on architectures that bear more similarity to the primate visual hierarchy, such as deep convolutional networks~\cite{kriegeskorte2015deep}, datasets of naturalistic images such as ImageNet~\cite{russakovsky2015imagenet}, and general naturalistic tasks such as visual question answering~\cite{agrawal2017vqa}.
This will allow us to assess whether the functional advantage provided by early modulation holds true in a more realistic scenario, and whether the resulting modulation schemes resemble those observed in the early visual areas of the primate brain.

Finally, a key aspect of our approach is the fact that the network is constantly operating in a task-dependent manner.
Most previous approaches to task-dependent modulation have assumed the presence of an underlying task-free representation on which the modulation operates (for example, in the case of \citeA{lindsay2018biological} this corresponds to a network pre-trained on object recognition).
Providing the network with task cues during the training phase, on the other hand, has been used in the field of continual learning~\cite{cheung2019superposition, masse2018alleviating,yang2019task}, and according to one influential theory in neuroscience, the interplay between sparse, context-specific information encoded by the hippocampus and shared structural information in the neo-cortex is crucial for learning new tasks without overwriting previous ones~\cite{kumaran2016learning}. To our knowledge, the question of how the task-based modulations observed in visual cortex might be learned has not been explicitly addressed in previous literature.
On the one hand, it is possible that a context-free representation is learned first, possibly through unsupervised learning, and then modulated upon.
On the other, learning of representations and task modulations might interact at all stages, allowing the representations to be optimised for the type of modulations they are subject to.
Whether one scheme or the other constitutes a better explanation for the modulations observed in biological visual systems is an important direction for future research.

\section{Acknowledgments}

This work was funded by the European Research Council (ERC) under the European Union's Horizon 2020 research and innovation program (Grant Agreement No. $725970$). This manuscript reflects only the authors' view, and the agency is not responsible for any use that may be made of the information it contains.

\bibliographystyle{apacite}

\setlength{\bibleftmargin}{.125in}
\setlength{\bibindent}{-\bibleftmargin}

\bibliography{ccn_style}

\begin{thebibliography}{}

\bibitem [\protect \citeauthoryear {%
Agrawal%
\ \protect \BOthers {.}}{%
Agrawal%
\ \protect \BOthers {.}}{%
{\protect \APACyear {2017}}%
}]{%
agrawal2017vqa}
\APACinsertmetastar {%
agrawal2017vqa}%
\begin{APACrefauthors}%
Agrawal, A.%
, Lu, J.%
, Antol, S.%
, Mitchell, M.%
, Zitnick, C\BPBI L.%
, Parikh, D.%
\BCBL {}\ \BBA {} Batra, D.%
\end{APACrefauthors}%
\unskip\
\newblock
\APACrefYearMonthDay{2017}{}{}.
\newblock
{\BBOQ}\APACrefatitle {Vqa: Visual question answering} {Vqa: Visual question
  answering}.{\BBCQ}
\newblock
\APACjournalVolNumPages{{International Journal of Computer
  Vision}}{123}{1}{4--31}.
\PrintBackRefs{\CurrentBib}

\bibitem [\protect \citeauthoryear {%
Boynton%
}{%
Boynton%
}{%
{\protect \APACyear {2009}}%
}]{%
boynton2009framework}
\APACinsertmetastar {%
boynton2009framework}%
\begin{APACrefauthors}%
Boynton, G\BPBI M.%
\end{APACrefauthors}%
\unskip\
\newblock
\APACrefYearMonthDay{2009}{}{}.
\newblock
{\BBOQ}\APACrefatitle {A framework for describing the effects of attention on
  visual responses} {A framework for describing the effects of attention on
  visual responses}.{\BBCQ}
\newblock
\APACjournalVolNumPages{{Vision Research}}{49}{10}{1129--1143}.
\PrintBackRefs{\CurrentBib}

\bibitem [\protect \citeauthoryear {%
Carrasco%
}{%
Carrasco%
}{%
{\protect \APACyear {2011}}%
}]{%
carrasco2011visual}
\APACinsertmetastar {%
carrasco2011visual}%
\begin{APACrefauthors}%
Carrasco, M.%
\end{APACrefauthors}%
\unskip\
\newblock
\APACrefYearMonthDay{2011}{}{}.
\newblock
{\BBOQ}\APACrefatitle {Visual attention: The past 25 years} {Visual attention:
  The past 25 years}.{\BBCQ}
\newblock
\APACjournalVolNumPages{{Vision Research}}{51}{13}{1484--1525}.
\PrintBackRefs{\CurrentBib}

\bibitem [\protect \citeauthoryear {%
Cheung%
, Terekhov%
, Chen%
, Agrawal%
\BCBL {}\ \BBA {} Olshausen%
}{%
Cheung%
\ \protect \BOthers {.}}{%
{\protect \APACyear {2019}}%
}]{%
cheung2019superposition}
\APACinsertmetastar {%
cheung2019superposition}%
\begin{APACrefauthors}%
Cheung, B.%
, Terekhov, A.%
, Chen, Y.%
, Agrawal, P.%
\BCBL {}\ \BBA {} Olshausen, B.%
\end{APACrefauthors}%
\unskip\
\newblock
\APACrefYearMonthDay{2019}{}{}.
\newblock
{\BBOQ}\APACrefatitle {Superposition of many models into one} {Superposition of
  many models into one}.{\BBCQ}
\newblock
\APACjournalVolNumPages{{arXiv preprint arXiv:1902.05522}}{}{}{}.
\PrintBackRefs{\CurrentBib}

\bibitem [\protect \citeauthoryear {%
Gilbert%
\ \BBA {} Li%
}{%
Gilbert%
\ \BBA {} Li%
}{%
{\protect \APACyear {2013}}%
}]{%
gilbert2013top}
\APACinsertmetastar {%
gilbert2013top}%
\begin{APACrefauthors}%
Gilbert, C\BPBI D.%
\BCBT {}\ \BBA {} Li, W.%
\end{APACrefauthors}%
\unskip\
\newblock
\APACrefYearMonthDay{2013}{}{}.
\newblock
{\BBOQ}\APACrefatitle {Top-down influences on visual processing} {Top-down
  influences on visual processing}.{\BBCQ}
\newblock
\APACjournalVolNumPages{{Nature Reviews Neuroscience}}{14}{5}{350--363}.
\PrintBackRefs{\CurrentBib}

\bibitem [\protect \citeauthoryear {%
Jehee%
, Brady%
\BCBL {}\ \BBA {} Tong%
}{%
Jehee%
\ \protect \BOthers {.}}{%
{\protect \APACyear {2011}}%
}]{%
jehee2011attention}
\APACinsertmetastar {%
jehee2011attention}%
\begin{APACrefauthors}%
Jehee, J\BPBI F.%
, Brady, D\BPBI K.%
\BCBL {}\ \BBA {} Tong, F.%
\end{APACrefauthors}%
\unskip\
\newblock
\APACrefYearMonthDay{2011}{}{}.
\newblock
{\BBOQ}\APACrefatitle {Attention improves encoding of task-relevant features in
  the human visual cortex} {Attention improves encoding of task-relevant
  features in the human visual cortex}.{\BBCQ}
\newblock
\APACjournalVolNumPages{{Journal of Neuroscience}}{31}{22}{8210--8219}.
\PrintBackRefs{\CurrentBib}

\bibitem [\protect \citeauthoryear {%
Kingma%
\ \BBA {} Ba%
}{%
Kingma%
\ \BBA {} Ba%
}{%
{\protect \APACyear {2014}}%
}]{%
kingma2014adam}
\APACinsertmetastar {%
kingma2014adam}%
\begin{APACrefauthors}%
Kingma, D\BPBI P.%
\BCBT {}\ \BBA {} Ba, J.%
\end{APACrefauthors}%
\unskip\
\newblock
\APACrefYearMonthDay{2014}{}{}.
\newblock
{\BBOQ}\APACrefatitle {Adam: A method for stochastic optimization} {Adam: A
  method for stochastic optimization}.{\BBCQ}
\newblock
\APACjournalVolNumPages{arXiv preprint arXiv:1412.6980}{}{}{}.
\PrintBackRefs{\CurrentBib}

\bibitem [\protect \citeauthoryear {%
Kirkpatrick%
\ \protect \BOthers {.}}{%
Kirkpatrick%
\ \protect \BOthers {.}}{%
{\protect \APACyear {2017}}%
}]{%
kirkpatrick2017overcoming}
\APACinsertmetastar {%
kirkpatrick2017overcoming}%
\begin{APACrefauthors}%
Kirkpatrick, J.%
, Pascanu, R.%
, Rabinowitz, N.%
, Veness, J.%
, Desjardins, G.%
, Rusu, A\BPBI A.%
\BDBL {}others%
\end{APACrefauthors}%
\unskip\
\newblock
\APACrefYearMonthDay{2017}{}{}.
\newblock
{\BBOQ}\APACrefatitle {Overcoming catastrophic forgetting in neural networks}
  {Overcoming catastrophic forgetting in neural networks}.{\BBCQ}
\newblock
\APACjournalVolNumPages{{Proceedings of the National Academy of
  Sciences}}{114}{13}{3521--3526}.
\PrintBackRefs{\CurrentBib}

\bibitem [\protect \citeauthoryear {%
Kriegeskorte%
}{%
Kriegeskorte%
}{%
{\protect \APACyear {2015}}%
}]{%
kriegeskorte2015deep}
\APACinsertmetastar {%
kriegeskorte2015deep}%
\begin{APACrefauthors}%
Kriegeskorte, N.%
\end{APACrefauthors}%
\unskip\
\newblock
\APACrefYearMonthDay{2015}{}{}.
\newblock
{\BBOQ}\APACrefatitle {Deep neural networks: a new framework for modeling
  biological vision and brain information processing} {Deep neural networks: a
  new framework for modeling biological vision and brain information
  processing}.{\BBCQ}
\newblock
\APACjournalVolNumPages{{Annual Review of Vision Science}}{1}{}{417--446}.
\PrintBackRefs{\CurrentBib}

\bibitem [\protect \citeauthoryear {%
Kumaran%
, Hassabis%
\BCBL {}\ \BBA {} McClelland%
}{%
Kumaran%
\ \protect \BOthers {.}}{%
{\protect \APACyear {2016}}%
}]{%
kumaran2016learning}
\APACinsertmetastar {%
kumaran2016learning}%
\begin{APACrefauthors}%
Kumaran, D.%
, Hassabis, D.%
\BCBL {}\ \BBA {} McClelland, J\BPBI L.%
\end{APACrefauthors}%
\unskip\
\newblock
\APACrefYearMonthDay{2016}{}{}.
\newblock
{\BBOQ}\APACrefatitle {What learning systems do intelligent agents need?
  Complementary learning systems theory updated} {What learning systems do
  intelligent agents need? complementary learning systems theory
  updated}.{\BBCQ}
\newblock
\APACjournalVolNumPages{{Trends in Cognitive Sciences}}{20}{7}{512--534}.
\PrintBackRefs{\CurrentBib}

\bibitem [\protect \citeauthoryear {%
LeCun%
, Bottou%
, Bengio%
\BCBL {}\ \BBA {} Haffner%
}{%
LeCun%
\ \protect \BOthers {.}}{%
{\protect \APACyear {1998}}%
}]{%
lecun1998gradient}
\APACinsertmetastar {%
lecun1998gradient}%
\begin{APACrefauthors}%
LeCun, Y.%
, Bottou, L.%
, Bengio, Y.%
\BCBL {}\ \BBA {} Haffner, P.%
\end{APACrefauthors}%
\unskip\
\newblock
\APACrefYearMonthDay{1998}{}{}.
\newblock
{\BBOQ}\APACrefatitle {Gradient-based learning applied to document recognition}
  {Gradient-based learning applied to document recognition}.{\BBCQ}
\newblock
\APACjournalVolNumPages{{Proceedings of the IEEE}}{86}{11}{2278--2324}.
\PrintBackRefs{\CurrentBib}

\bibitem [\protect \citeauthoryear {%
Lindsay%
\ \BBA {} Miller%
}{%
Lindsay%
\ \BBA {} Miller%
}{%
{\protect \APACyear {2018}}%
}]{%
lindsay2018biological}
\APACinsertmetastar {%
lindsay2018biological}%
\begin{APACrefauthors}%
Lindsay, G\BPBI W.%
\BCBT {}\ \BBA {} Miller, K\BPBI D.%
\end{APACrefauthors}%
\unskip\
\newblock
\APACrefYearMonthDay{2018}{}{}.
\newblock
{\BBOQ}\APACrefatitle {How biological attention mechanisms improve task
  performance in a large-scale visual system model} {How biological attention
  mechanisms improve task performance in a large-scale visual system
  model}.{\BBCQ}
\newblock
\APACjournalVolNumPages{{eLife}}{7}{}{e38105}.
\PrintBackRefs{\CurrentBib}

\bibitem [\protect \citeauthoryear {%
Ling%
, Liu%
\BCBL {}\ \BBA {} Carrasco%
}{%
Ling%
\ \protect \BOthers {.}}{%
{\protect \APACyear {2009}}%
}]{%
ling2009spatial}
\APACinsertmetastar {%
ling2009spatial}%
\begin{APACrefauthors}%
Ling, S.%
, Liu, T.%
\BCBL {}\ \BBA {} Carrasco, M.%
\end{APACrefauthors}%
\unskip\
\newblock
\APACrefYearMonthDay{2009}{}{}.
\newblock
{\BBOQ}\APACrefatitle {How spatial and feature-based attention affect the gain
  and tuning of population responses} {How spatial and feature-based attention
  affect the gain and tuning of population responses}.{\BBCQ}
\newblock
\APACjournalVolNumPages{{Vision Research}}{49}{10}{1194--1204}.
\PrintBackRefs{\CurrentBib}

\bibitem [\protect \citeauthoryear {%
Mante%
, Sussillo%
, Shenoy%
\BCBL {}\ \BBA {} Newsome%
}{%
Mante%
\ \protect \BOthers {.}}{%
{\protect \APACyear {2013}}%
}]{%
mante2013context}
\APACinsertmetastar {%
mante2013context}%
\begin{APACrefauthors}%
Mante, V.%
, Sussillo, D.%
, Shenoy, K\BPBI V.%
\BCBL {}\ \BBA {} Newsome, W\BPBI T.%
\end{APACrefauthors}%
\unskip\
\newblock
\APACrefYearMonthDay{2013}{}{}.
\newblock
{\BBOQ}\APACrefatitle {Context-dependent computation by recurrent dynamics in
  prefrontal cortex} {Context-dependent computation by recurrent dynamics in
  prefrontal cortex}.{\BBCQ}
\newblock
\APACjournalVolNumPages{{Nature}}{503}{7474}{78}.
\PrintBackRefs{\CurrentBib}

\bibitem [\protect \citeauthoryear {%
Masse%
, Grant%
\BCBL {}\ \BBA {} Freedman%
}{%
Masse%
\ \protect \BOthers {.}}{%
{\protect \APACyear {2018}}%
}]{%
masse2018alleviating}
\APACinsertmetastar {%
masse2018alleviating}%
\begin{APACrefauthors}%
Masse, N\BPBI Y.%
, Grant, G\BPBI D.%
\BCBL {}\ \BBA {} Freedman, D\BPBI J.%
\end{APACrefauthors}%
\unskip\
\newblock
\APACrefYearMonthDay{2018}{}{}.
\newblock
{\BBOQ}\APACrefatitle {Alleviating catastrophic forgetting using
  context-dependent gating and synaptic stabilization} {Alleviating
  catastrophic forgetting using context-dependent gating and synaptic
  stabilization}.{\BBCQ}
\newblock
\APACjournalVolNumPages{{Proceedings of the National Academy of
  Sciences}}{115}{44}{E10467--E10475}.
\PrintBackRefs{\CurrentBib}

\bibitem [\protect \citeauthoryear {%
Maunsell%
\ \BBA {} Treue%
}{%
Maunsell%
\ \BBA {} Treue%
}{%
{\protect \APACyear {2006}}%
}]{%
maunsell2006feature}
\APACinsertmetastar {%
maunsell2006feature}%
\begin{APACrefauthors}%
Maunsell, J\BPBI H.%
\BCBT {}\ \BBA {} Treue, S.%
\end{APACrefauthors}%
\unskip\
\newblock
\APACrefYearMonthDay{2006}{}{}.
\newblock
{\BBOQ}\APACrefatitle {Feature-based attention in visual cortex} {Feature-based
  attention in visual cortex}.{\BBCQ}
\newblock
\APACjournalVolNumPages{{Trends in Neurosciences}}{29}{6}{317--322}.
\PrintBackRefs{\CurrentBib}

\bibitem [\protect \citeauthoryear {%
McNemar%
}{%
McNemar%
}{%
{\protect \APACyear {1947}}%
}]{%
mcnemar1947note}
\APACinsertmetastar {%
mcnemar1947note}%
\begin{APACrefauthors}%
McNemar, Q.%
\end{APACrefauthors}%
\unskip\
\newblock
\APACrefYearMonthDay{1947}{}{}.
\newblock
{\BBOQ}\APACrefatitle {Note on the sampling error of the difference between
  correlated proportions or percentages} {Note on the sampling error of the
  difference between correlated proportions or percentages}.{\BBCQ}
\newblock
\APACjournalVolNumPages{Psychometrika}{12}{2}{153--157}.
\PrintBackRefs{\CurrentBib}

\bibitem [\protect \citeauthoryear {%
Rosenblatt%
}{%
Rosenblatt%
}{%
{\protect \APACyear {1957}}%
}]{%
rosenblatt1957perceptron}
\APACinsertmetastar {%
rosenblatt1957perceptron}%
\begin{APACrefauthors}%
Rosenblatt, F.%
\end{APACrefauthors}%
\unskip\
\newblock
\APACrefYear{1957}.
\newblock
\APACrefbtitle {The perceptron, a perceiving and recognizing automaton} {The
  perceptron, a perceiving and recognizing automaton}.
\newblock
\APACaddressPublisher{}{Cornell Aeronautical Laboratory}.
\PrintBackRefs{\CurrentBib}

\bibitem [\protect \citeauthoryear {%
Rosenfeld%
, Biparva%
\BCBL {}\ \BBA {} Tsotsos%
}{%
Rosenfeld%
\ \protect \BOthers {.}}{%
{\protect \APACyear {2018}}%
}]{%
rosenfeld2018priming}
\APACinsertmetastar {%
rosenfeld2018priming}%
\begin{APACrefauthors}%
Rosenfeld, A.%
, Biparva, M.%
\BCBL {}\ \BBA {} Tsotsos, J\BPBI K.%
\end{APACrefauthors}%
\unskip\
\newblock
\APACrefYearMonthDay{2018}{}{}.
\newblock
{\BBOQ}\APACrefatitle {Priming Neural Networks} {Priming neural
  networks}.{\BBCQ}
\newblock
\BIn{} \APACrefbtitle {{Proceedings of the IEEE Conference on Computer Vision
  and Pattern Recognition Workshops}} {{Proceedings of the IEEE Conference on
  Computer Vision and Pattern Recognition Workshops}}\ (\BPGS\ 2011--2020).
\PrintBackRefs{\CurrentBib}

\bibitem [\protect \citeauthoryear {%
Russakovsky%
\ \protect \BOthers {.}}{%
Russakovsky%
\ \protect \BOthers {.}}{%
{\protect \APACyear {2015}}%
}]{%
russakovsky2015imagenet}
\APACinsertmetastar {%
russakovsky2015imagenet}%
\begin{APACrefauthors}%
Russakovsky, O.%
, Deng, J.%
, Su, H.%
, Krause, J.%
, Satheesh, S.%
, Ma, S.%
\BDBL {}others%
\end{APACrefauthors}%
\unskip\
\newblock
\APACrefYearMonthDay{2015}{}{}.
\newblock
{\BBOQ}\APACrefatitle {Imagenet large scale visual recognition challenge}
  {Imagenet large scale visual recognition challenge}.{\BBCQ}
\newblock
\APACjournalVolNumPages{{International Journal of Computer
  Vision}}{115}{3}{211--252}.
\PrintBackRefs{\CurrentBib}

\bibitem [\protect \citeauthoryear {%
Thorat%
, van Gerven%
\BCBL {}\ \BBA {} Peelen%
}{%
Thorat%
\ \protect \BOthers {.}}{%
{\protect \APACyear {2018}}%
}]{%
thorat2018functional}
\APACinsertmetastar {%
thorat2018functional}%
\begin{APACrefauthors}%
Thorat, S.%
, van Gerven, M.%
\BCBL {}\ \BBA {} Peelen, M.%
\end{APACrefauthors}%
\unskip\
\newblock
\APACrefYearMonthDay{2018}{}{}.
\newblock
{\BBOQ}\APACrefatitle {The functional role of cue-driven feature-based feedback
  in object recognition} {The functional role of cue-driven feature-based
  feedback in object recognition}.{\BBCQ}
\newblock
\BIn{} \APACrefbtitle {{Conference on Cognitive Computational Neuroscience,
  {CCN} 2018}} {{Conference on Cognitive Computational Neuroscience, {CCN}
  2018}}\ (\BPGS\ 1--4).
\PrintBackRefs{\CurrentBib}

\bibitem [\protect \citeauthoryear {%
Yang%
, Joglekar%
, Song%
, Newsome%
\BCBL {}\ \BBA {} Wang%
}{%
Yang%
\ \protect \BOthers {.}}{%
{\protect \APACyear {2019}}%
}]{%
yang2019task}
\APACinsertmetastar {%
yang2019task}%
\begin{APACrefauthors}%
Yang, G\BPBI R.%
, Joglekar, M\BPBI R.%
, Song, H\BPBI F.%
, Newsome, W\BPBI T.%
\BCBL {}\ \BBA {} Wang, X\BHBI J.%
\end{APACrefauthors}%
\unskip\
\newblock
\APACrefYearMonthDay{2019}{}{}.
\newblock
{\BBOQ}\APACrefatitle {Task representations in neural networks trained to
  perform many cognitive tasks} {Task representations in neural networks
  trained to perform many cognitive tasks}.{\BBCQ}
\newblock
\APACjournalVolNumPages{{Nature Neuroscience}}{22}{2}{297}.
\PrintBackRefs{\CurrentBib}

\end{thebibliography}

\section{Appendix}

Here we present the results from further analyses performed to address reviewer comments, post-acceptance into the 2019 Conference on Cognitive Computational Neuroscience.

\subsection{Dependence of the results on parameter expansion}

When early modulation is performed in addition to late modulation, the addition of parameters is constant across the different networks with varying number of hidden neurons. However, the relative increase in the number of parameters is higher in the network with lower number of neurons in the hidden layer. This network, with lower number of hidden neurons, also benefits most from the addition of early modulation, as observed in Figure~\ref{fig:fig3}. So, are the observations in Figure~\ref{fig:fig3} an effect of simple proportionate parameter expansion? Two additional analyses show that this is not the case.

\subsubsection{Matching the proportionate increase in the number of parameters across networks}

\begin{figure}[!ht]
\centering
\includegraphics[width=0.4\textwidth]{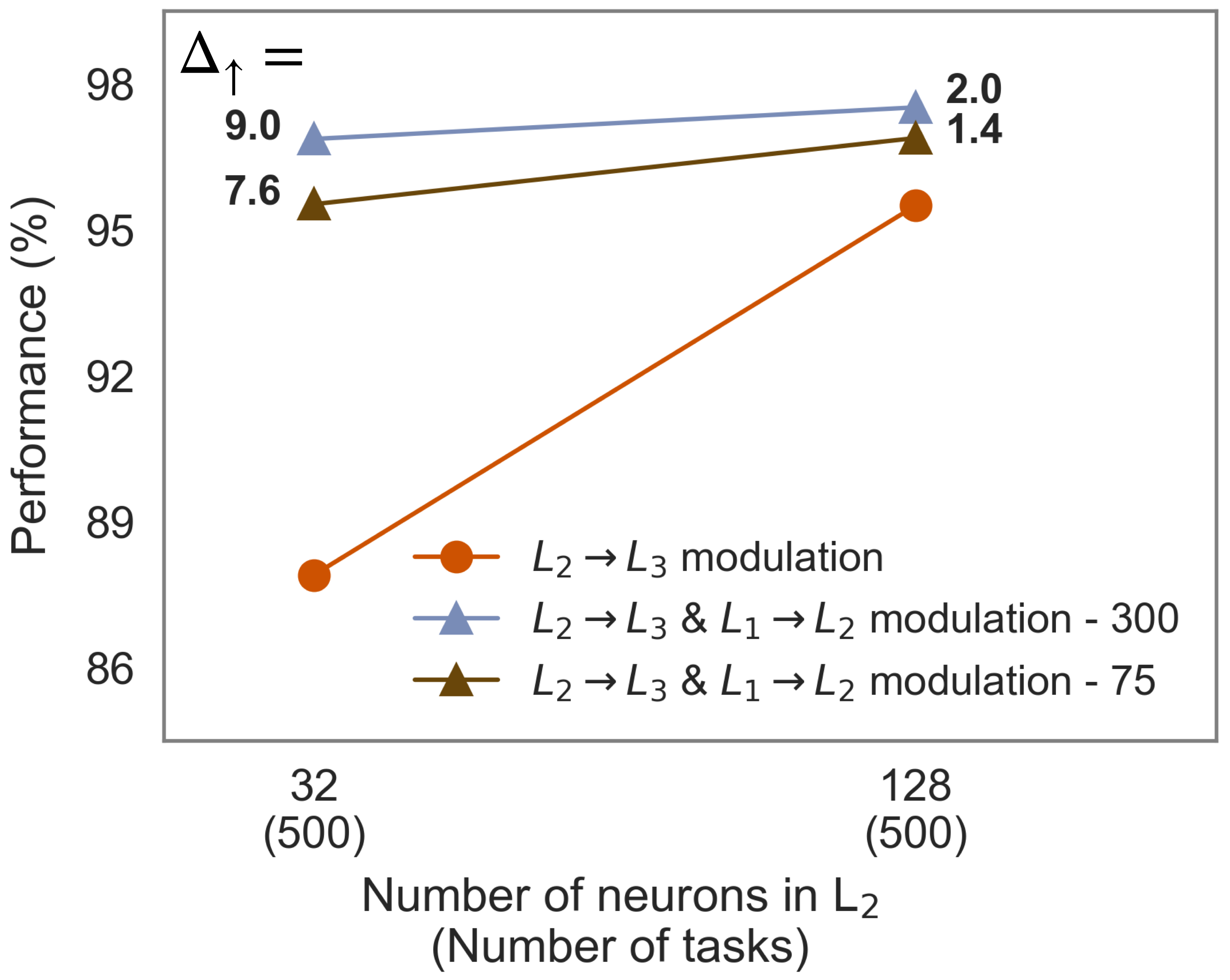}
\caption{The effectiveness of task-based \textit{gain} modulation (quantified by the performance boost, $\Delta_{\uparrow}$) of early neural processing ($L_1\rightarrow L_2$), as a function of the number of neurons in $L_2$ and the number of neurons in the hidden layer of early modulation. Reducing the number of hidden neurons to $75$ does not reduce the performance boost substantially. This observation suggests that the trends observed in Figure~\ref{fig:fig3} are not simply a consequence of a disproportionate increase in the number of parameters in the different networks.}
\label{fig:fig4}
\end{figure}

To match the proportionate increase in parameters, we introduced an additional layer between the task cue $c$ and the gain modulation $g_1$ to the first (input) layer of the network. No bias modulation was included in these networks as we found it did not aid gain modulation. We set the number of hidden neurons in this modulation network to $300$. We wanted to match the proportionate increase in parameters accompanying the addition of early gain modulation to late gain modulation, between two networks with $32$ and $128$ neurons in $L_2$ respectively. To accomplish this matching, in the case of the network with $32$ neurons the hidden layer of the modulation needs to contain approximately $75$ neurons. As seen in Figure~\ref{fig:fig4}, reducing the number of hidden neurons in early modulation to $75$, to match the proportionate increase in parameters between the two networks with $32$ and $128$ neurons in $L_2$, does not substantially reduce the increase in performance provided by task-based modulation of early neural processing.

\subsubsection{Increasing the task difficulty to increase capacity constraints}

\begin{figure}[!ht]
\centering
\includegraphics[width=0.4\textwidth]{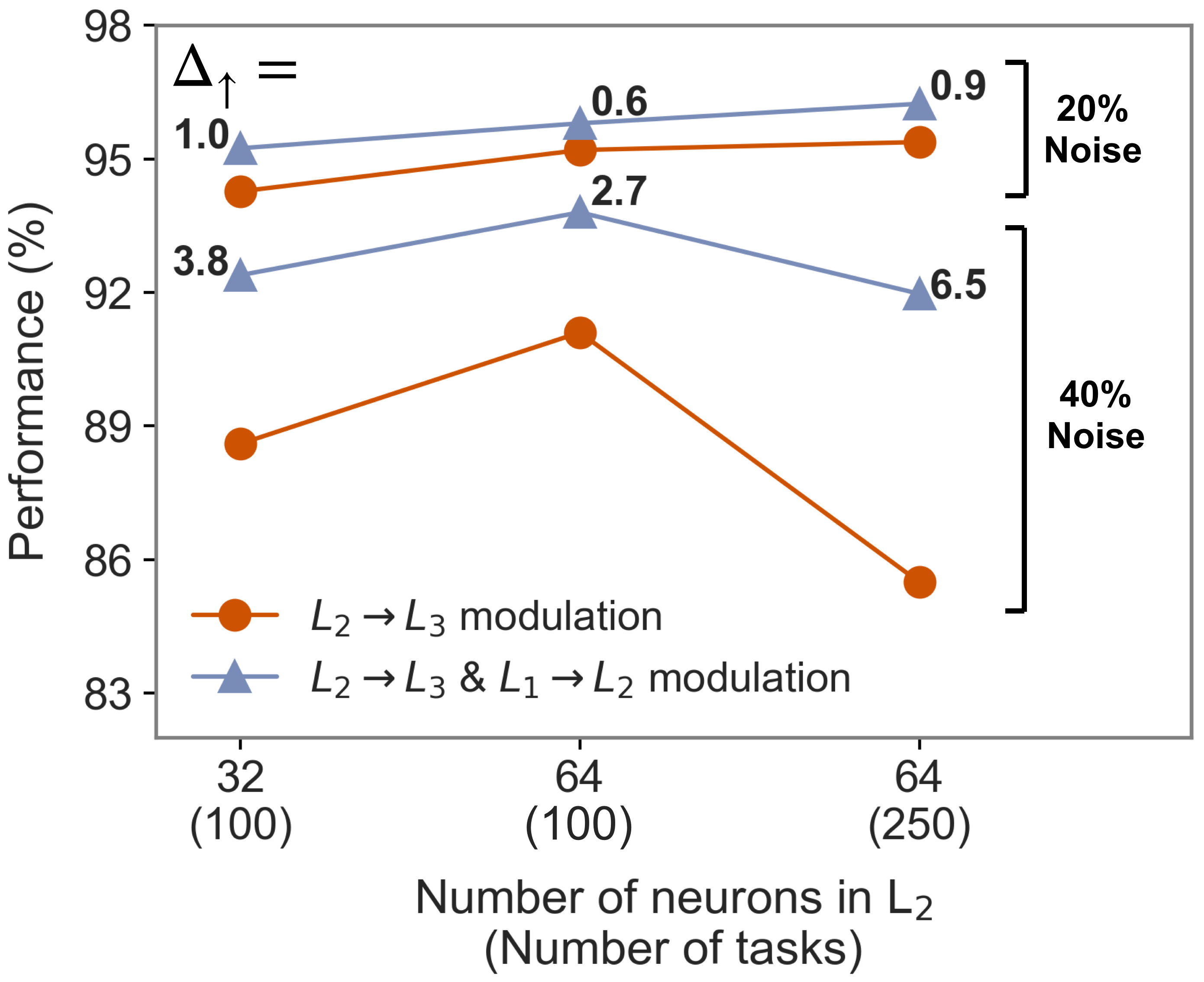}
\caption{The effect of increasing task difficulty on the additive effectiveness of early modulation. The performance boost ($\Delta_\uparrow$) increases when noise in the stimuli is increased making each task more difficult. The performance boost in the high noise case also decreases with increasing number of neurons in $L_2$, and increases with increasing number of tasks to be performed, echoing the trends in Figure~\ref{fig:fig3}.}
\label{fig:fig5}
\end{figure}

In addition to manipulating the number of hidden neurons in the network and the number of tasks to be performed, the difficulty of the tasks could also contribute to capacity limits. If we increase the noise in the stimuli from $20\%$ (which is the case in the main analysis; additive uniform random noise) to $40\%$, the performance boost increases as seen in Figure~\ref{fig:fig5}. As the number of parameters stays the same across the change in the level of noise, this result also suggests that the trends observed in Figure~\ref{fig:fig3} are not simply a consequence of simple proportionate parameter expansion.

%\begin{table}[!ht]
%\begin{center} 
%\begin{tabular}{llllll} 
%\hline
%\#N    &  \#T & noise & late only & early+late & $\boldsymbol{\Delta}_\uparrow$ \\
%&& level & performance & performance \\
%\hline
%32        &   100 &   20\% &   94.3\% & 95.2\% & 0.9\% \\
%32        &   100 &   40\% &   88.4\% & 92.2\% & 3.8\% \\
%64        &   100 &   20\% &   95.2\% & 95.8\% & 0.6\% \\
%64        &   100 &   40\% &   91.1\% & 93.8\% & 2.7\% \\
%64        &   250 &   20\% &   95.4\% & 96.2\% & 0.8\% \\
%64        &   250 &   40\% &   85.7\% & 91.9\% & 6.2\% \\
%\hline
%\end{tabular} 
%\caption{The effect of increasing task difficulty on the additive effectiveness of early modulation. The %performance boost ($\Delta_\uparrow$) increases when noise in the stimuli is increased making each task more %difficult.}
%\label{tab:sample-table} 
%\end{center} 
%\end{table}

\subsection{Comparison with \citeA{cheung2019superposition}}

The task-based modulation scheme in our work was adapted from~\citeA{cheung2019superposition}. However, there are major differences between our work and \citeA{cheung2019superposition}. Their work proposes a solution to the problem of catastrophic forgetting in neural networks when faced with a continual stream of tasks. In our work, the tasks are interleaved. They used random task-based modulations, while in our work the task-based modulations are learned to optimise detection performance. If two tasks are similar, the task-based modulations should also be similar and this relationship can be learned by the networks in our work. 
We wanted to assess if learning the modulations instead of fixing them randomly, as in \citeA{cheung2019superposition}, aids performance on the detection tasks here. To do so, we trained a network with $32$ neurons in $L_2$ on $500$ tasks, with or without training the bias and gain modulations (early and late modulation) which are initialised with random binary vectors as in \citeA{cheung2019superposition}. When the modulation schemes can be learned, the performance of the network was $94.0\%$, while when the scheme was fixed as random binary vectors, the performance was $71.1\%$, confirming the idea that task-based modulations can account for similarities across tasks. How such task similarities could be included in learning task-based modulations in a continual learning setting is an open question and beyond the scope of our work.

\subsection{Robustness of presented effects}

To assess if the differences in the performance ($\Delta_\uparrow$) of the networks mentioned in Figure~\ref{fig:fig3} are robust, we used McNemar's test~\cite{mcnemar1947note}, a statistical test used on paired nominal data. This test compares the quantities of examples where the decisions of the two networks being compared differ. If one network misclassifies examples the other network classifies correctly more often (say $b$ examples) than the other way around (say $c$ examples), the test statistic ($\chi^2 = (b-c)^2/(b+c)$) is higher. The test statistic has a chi-squared distribution with $1$ degree of freedom. 

For each network (varying on the number of hidden neurons and the number of tasks performed), we compared the outputs when only late modulation was active and when both early and late modulation (global modulation) were active. Across all the comparisons, the $\chi^2$ values were above $15$ (which corresponds to $p=10^{-4}$). So, all the differences ($\Delta_\uparrow$) in Figure~\ref{fig:fig3} correspond to robust differences in the performance of the corresponding networks.

\subsection{Additional observations about the behaviour of the trained neural networks}

Below we clarify the training setup and present an additional observation about the effectiveness of early modulation.

\subsubsection{The network mainly performs permutation discrimination}

In training the networks, each batch contained $50$ examples corresponding to the cue (example cue: $5$ in permutation $10$, corresponding to task $95$) and $50$ examples corresponding to every other task (the invalid case, where the network outputs 'No'). As the invalid cases are drawn randomly, the probability that the same digit as the valid case ($5$) would be included is $1/10$. The probability that the same permutation as the valid case (permutation $10$) would be included is $1/50$ (in the case of $50$ permutations). In this setting, the network might thus mainly perform permutation discrimination rather than digit discrimination. To show that this was indeed the case, we probed the behaviour of a network with $32$ neurons in $L_2$ trained to perform $500$ tasks (with joint modulation). When we only included permutation-matched digits as invalid test examples, the performance was $60.9\%$, while when we only included digit-matched permutations as invalid test examples, the performance was $94.3\%$. This demonstrates that the high performance of the network (Figure~\ref{fig:fig3}) largely reflected permutation discrimination.

\subsubsection{Modulating late neural processing in addition to modulating early neural processing does not boost performance}

How well does modulating early neural processing alone perform? We trained a network with $32$ neurons in $L_2$ to perform $250$ or $500$ tasks, with only the modulation of early neural processing. The performance of this network was equal to the performance of the same network trained with joint modulation. This suggests that, in this setting, only modulating early neural processing is better than only modulating late neural processing, and that late modulation does not aid performance on top of early modulation.

Early modulation might be performing so well because the stimuli used might be distinguishable at the pixel-level. For example, as the network is mostly performing permutation discrimination, each permutation might have a characteristic spread of pixels that the first transformation could pick on and distinguish between the valid and invalid permutations. Switching to naturalistic stimuli might remove such low-level distinctions between stimuli, thereby not allowing the network to capitalise solely on early modulation.

In biological systems, the trade-off between wiring costs and task optimality might exist even if early modulation is always equal to or better than late modulation. This is because it might only be in the (capacity-limited) cases, where late modulation performs poorly, that early modulation is worth its wiring costs. Further work involving naturalistic stimuli would provide a better understanding of the nature of this trade-off.

\end{document}